\documentclass[twocolumn,aps,prl,floatfix,superscriptaddress]{revtex4-1}
\usepackage{amsmath,amssymb}
\usepackage{graphicx,float}
\usepackage{times,txfonts}
\usepackage{color}
\usepackage{multirow}
\usepackage{bbold}
\usepackage{color}
\usepackage{soul}
\usepackage{hyperref}
\usepackage{times,txfonts}
\usepackage{natbib}
\usepackage{nicefrac}
\usepackage{blindtext}
\usepackage[export]{adjustbox}
\usepackage{mathrsfs}
\usepackage{textcomp}

\newcommand{\abs}[1]{\left| #1 \right|}

\renewcommand{\epsilon}{\varepsilon}
\renewcommand{\phi}{\varphi}
\renewcommand{\mod}{\text{mod}}

\usepackage{sistyle}
\usepackage{soul}
\definecolor{lightblue}{RGB}{185,210,248}
\sethlcolor{lightblue}

\DeclareMathOperator{\sign}{sign}
\begin{document}
\title{Measuring an electron beam's orbital angular momentum spectrum}
\author{Vincenzo Grillo}
\affiliation{CNR-Istituto Nanoscienze, Centro S3, Via G Campi 213/a, I-41125 Modena, Italy}
\affiliation{CNR-IMEM Parco Area delle Scienze 37/A, I-43124 Parma, Italy}
\author{Amir H. Tavabi}
\affiliation{Ernst Ruska-Centre for Microscopy and Spectroscopy with Electrons, Forschungszentrum J\"ulich, J\"ulich 52425, Germany}
\author{Federico Venturi}
\affiliation{CNR-Istituto Nanoscienze, Centro S3, Via G Campi 213/a, I-41125 Modena, Italy}
\affiliation{Dipartimento FIM Universit\'a di Modenae Reggio Emilia, Via G Campi 213/a, I-41125 Modena, Italy}
\author{Hugo Larocque}
\affiliation{The Max Planck Centre for Extreme and Quantum Photonics, Department of Physics, University of Ottawa, 25 Templeton St., Ottawa, Ontario, K1N 6N5 Canada}
\author{Roberto Balboni}
\affiliation{CNR-IMM Bologna, Via P. Gobetti 101, 40129 Bologna, Italy}
\author{Gian Carlo Gazzadi}
\affiliation{CNR-Istituto Nanoscienze, Centro S3, Via G Campi 213/a, I-41125 Modena, Italy}
\author{Stefano Frabboni}
\affiliation{CNR-Istituto Nanoscienze, Centro S3, Via G Campi 213/a, I-41125 Modena, Italy}
\affiliation{Dipartimento FIM Universit\'a di Modenae Reggio Emilia, Via G Campi 213/a, I-41125 Modena, Italy}
\author{Peng-Han Lu}
\affiliation{Ernst Ruska-Centre for Microscopy and Spectroscopy with Electrons, Forschungszentrum J\"ulich, J\"ulich 52425, Germany}
\author{Erfan Mafakheri}
\affiliation{Dipartimento FIM Universit\'a di Modenae Reggio Emilia, Via G Campi 213/a, I-41125 Modena, Italy}
\author{Fr\'ed\'eric Bouchard}
\affiliation{The Max Planck Centre for Extreme and Quantum Photonics, Department of Physics, University of Ottawa, 25 Templeton St., Ottawa, Ontario, K1N 6N5 Canada}
\author{Rafal E. Dunin-Borkowski}
\affiliation{Ernst Ruska-Centre for Microscopy and Spectroscopy with Electrons, Forschungszentrum J\"ulich, J\"ulich 52425, Germany}
\author{Robert W. Boyd}
\affiliation{The Max Planck Centre for Extreme and Quantum Photonics, Department of Physics, University of Ottawa, 25 Templeton St., Ottawa, Ontario, K1N 6N5 Canada}
\affiliation{Institute of Optics, University of Rochester, Rochester, New York, 14627, USA}
\author{Martin P. J. Lavery}
\affiliation{School of Physics and Astronomy, Glasgow University, Glasgow, G12 8QQ, Scotland, UK}
\author{Miles J. Padgett}
\affiliation{School of Physics and Astronomy, Glasgow University, Glasgow, G12 8QQ, Scotland, UK}
\author{Ebrahim Karimi}
\affiliation{The Max Planck Centre for Extreme and Quantum Photonics, Department of Physics, University of Ottawa, 25 Templeton St., Ottawa, Ontario, K1N 6N5 Canada}
\affiliation{Department of Physics, Institute for Advanced Studies in Basic Sciences, 45137-66731 Zanjan, Iran}
\maketitle
\noindent\textbf{Quantum complementarity states that particles, e.g. electrons, can exhibit wave-like properties such as diffraction and interference upon propagation. \textit{Electron waves} defined by a helical wavefront are referred to as twisted electrons~\cite{uchida:10,verbeeck:10,mcmorran:11}. These electrons are also characterised by a quantized and unbounded magnetic dipole moment parallel to their propagation direction, as they possess a net charge of $-|e|$~\cite{bliokh:07}. When interacting with magnetic materials, the wavefunctions of twisted electrons are inherently modified~\cite{lloyd:12b,schattschneider:14a,asenjo:14}. Such variations therefore motivate the need to analyze electron wavefunctions, especially their wavefronts, in order to obtain information regarding the material's structure~\cite{harris:15}. Here, we propose, design, and demonstrate the performance of a device for measuring an electron's azimuthal wavefunction, i.e. its orbital angular momentum (OAM) content. Our device consists of nanoscale holograms designed to introduce astigmatism onto the electron wavefunctions and spatially separate its phase components. We sort pure and superposition OAM states of electrons ranging within OAM values of $-10$ and $10$. We employ the device to analyze the OAM spectrum of electrons having been affected by a micron-scale magnetic dipole, thus establishing that, with a midfield optical configuration, our sorter can be an instrument for nano-scale magnetic spectroscopy.}

The helical wavefronts of twisted electron wavefunctions bestow additional mechanical and magnetic properties to their massive and charged nature. For instance, upon elastic interaction, these magnetic properties along with the opposite handedness of twisted electrons allow for probing magnetic chirality as well as magnetic dichroism~\cite{lloyd:12b,asenjo:14,schattschneider:14a}. The added \emph{unbounded} twisted motion to these charged particles is pertinent to the investigation of the nature of radiation~\cite{kaminer:16}, virtual forces and increasing the lifetime of unstable and metastable particles~\cite{kaminer:15}. From a more fundamental viewpoint, structured electrons also provide novel insights into the quantum nature of electromagnetic-matter interaction, e.g. realisation of Landau-Zeeman states~\cite{bliokh:12,schattschneider:14} and spin-to-orbit coupling~\cite{bliokh:11}. OAM-carrying electron waves can be generated through a variety of methods by directly interacting with the electrons' wavefronts. These processes rely on devices such as spiral phase plates~\cite{uchida:10}, amplitude and phase holograms~\cite{verbeeck:10,harvey:14,grillo:14}, cylindrical lens~\cite{schattschneider:12b} mode converters, or even electron microscope corrector lenses~\cite{clark:13}. Spin-to-orbit coupling has also been theoretically proposed as a method to add OAM onto electrons~\cite{karimi:12}. Other methods to do so exploit the magnetic properties of electrons, most notably by employing a magnetic needle simulating a magnetic monopole~\cite{beche:14}.

Reciprocally, devices used to generate twisted electrons can also be adapted to measure an electron's OAM content~\cite{guzzinati:14}. The most commonly employed of these devices relies on a series of projective phase flattening measurements allowing one to obtain the magnitude of each OAM component of a beam's spectrum~\cite{mair:01}. To perform such an analysis, an OAM component's wavefront is flattened, thus causing it to gain a Gaussian-like profile upon propagation~\cite{saitoh:13}. This profile allows it to be easily selected from the remaining parts of the beam and therefore to evaluate the intensity of this component. To obtain these relative intensities, each component needs to be coupled to the device's electron detection mechanism. However, this coupling process is biased toward electrons carrying lower absolute values of OAM thus introducing discrepancies in the measured spectrum~\cite{qassim:14}. Moreover, much like how different types of OAM carrying beams are generated using different devices, different devices are required to flatten the wavefronts of the beam's various OAM components. Therefore, though such an analysis of a beam's OAM content is efficient, it also requires a substantial number of elements to perform repetitive measurements. There are also other OAM measurement methods relying on interferometry~\cite{leach:02}, though they possess serious limitations such as a limited amount of information that can be extracted from obtained interference patterns or also by the need of an extremely stable interferometer. Interferometry is also of limited usefulness when analyzing the OAM content of inelastically scattered electrons due to their short coherence lengths.

A viable alternative to these methods is available in optics and relies on transforming an OAM-carrying photon's azimuthal phase variations into transverse phase gradients that can be spatially resolved and separated with a lensing element~\cite{berkhout:10,mirhosseini:13}. The device thus effectively behaves as an OAM spectrometer which could be of significant interest in electron optics and materials science as it would provide detailed information on a material's magnetic spectrum. 

The operation of our presented electron OAM spectrometer relies on a similar OAM separation process. Much like its optical counterpart, it is essentially based on unwrapping the azimuthal phase variations associated with an electron's OAM into variations over a Cartesian coordinate of the plane transverse to the electron's propagation. This effectively causes the electron's original helical wavefronts to become planar and inclined with respect to the beam's original direction of propagation. Namely, the degree to which these wavefronts are tilted will increase with the azimuthal variation of the electron's original phase profile, i.e. its OAM. As a result, focusing these ``unwrapped'' waves with a lensing element will cause electrons originally carrying different OAM values to focus at correspondingly separate lateral positions. By using this method, we are thus able to decompose an electron beam's OAM content by measuring the relative electron intensity at each of these possible focusing positions.

\begin{figure}[h]
	\begin{center}
	\includegraphics[width=1\columnwidth]{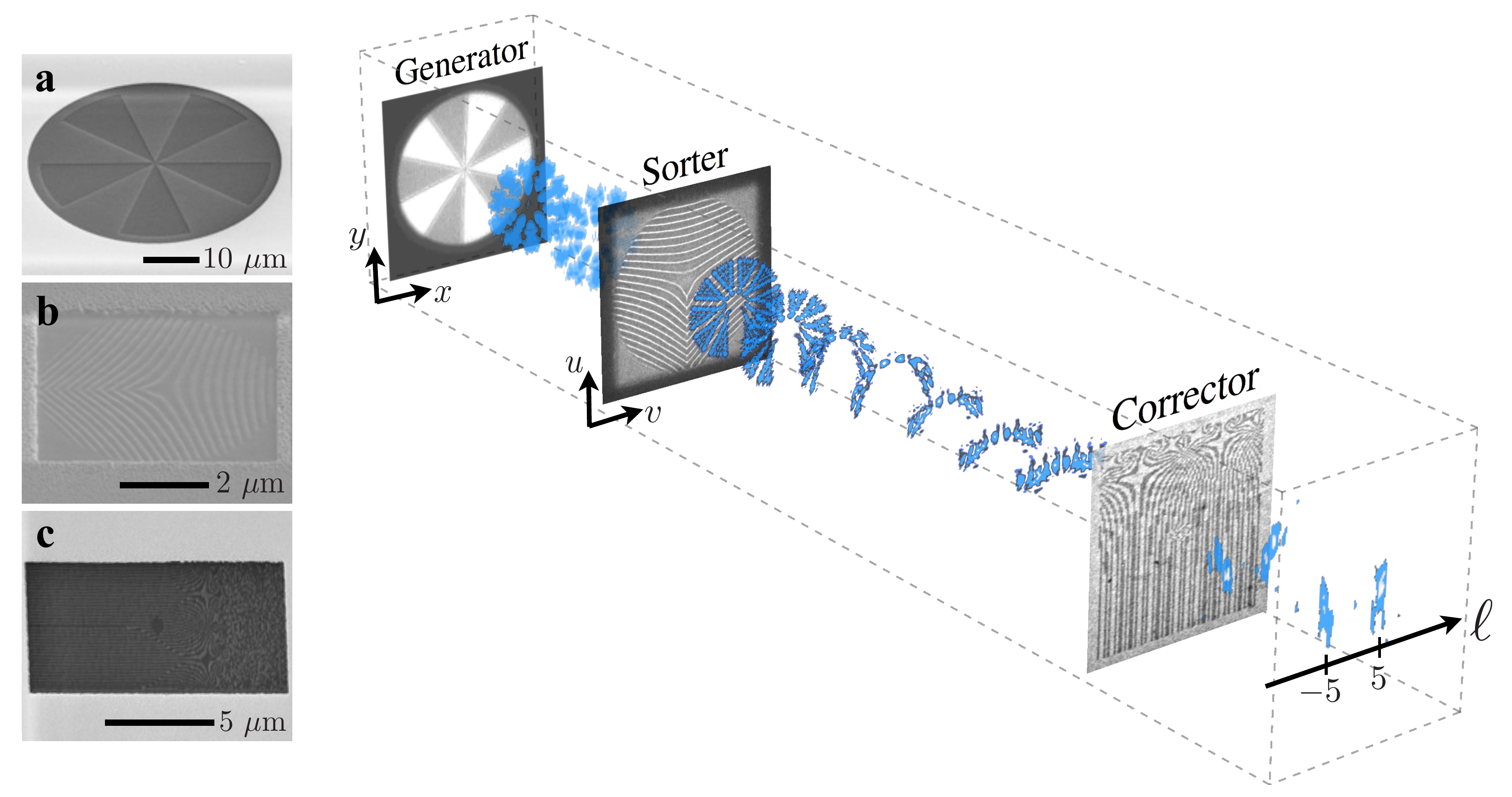}
	\caption[]{{\bf Schematics of the electron sorter depicting TEM images of the phase holograms.} These schematics also show an electron beam's experimental transverse intensity profile recorded at various planes in the sorting apparatus. A hologram in the sorter's generator plane, which corresponds to the electron microscope's condensor, produces an electron beam carrying OAM. In this particular case, the beam consists of a superposition of $\pm5$ OAM states. The beam then goes through a hologram in the apparatus' sorter plane, positioned at the microscope's sample holder, which performs the required conformal mapping $(x,y)\mapsto(u,v)$. Once the beam is unwrapped, it passes through a hologram in the sorter's corrector plane corresponding to the microscope's SAD aperture. This hologram brings corrections to any phase defects to the beam in order to stabilize its propagation through the rest of the sorter. At the sorter's output, the original beam's OAM content is spatially resolved on a screen and captured by a CCD camera. Scanning electron microscopy (SEM) images of the depicted holograms, the ones in the generator, sorter, and corrector planes, are shown in {\bf a}, {\bf b}, and {\bf c}, respectively.}
	\label{fig:setupOAM5}
	\end{center}
\end{figure}

The unwrapping process, as detailed in \cite{berkhout:10}, requires the beam's transverse profile in polar coordinates ($r,\phi$) to be mapped to Cartesian coordinates. Such a transformation can be achieved by means of a conformal mapping between the Cartesian coordinates of the initial beam's profile ($x,y$) and those of its final profile ($u,v$). This conformal mapping, $\Phi$, involves taking the logarithm of the beam's transverse wavefunction $\psi_\ell(r,\phi)=f(r)\exp{(i \ell \phi)}$ and scaling it with parameters $a$ and $b$, i.e. $\Phi(\psi_\ell)=a\ln{(\psi_\ell/b)}$.  Performing such a mapping will result in the following wavefunction
\begin{align}
\begin{split}
	\Phi(f(r)\exp{( i\ell \phi)}) &=  U+i\,V \\
	&= a \ln{(f(r)/b)}  + i\,a\ell\phi.
\end{split}
\end{align}
Such mappings are commonly conducted using phase elements performing a log-polar coordinate transformation satisfying $u=-a\ln{(r/b)}$ and $v=a\phi$. We adopt a similar diffractive approach to develop our sorter using two electron phase holograms (see Supplementary Materials for more details) as displayed in Fig.~\ref{fig:setupOAM5}.

The first of the holograms effectively maps the electron's azimuthal phase variations along a Cartesian coordinate while the second provides phase corrections to defects induced by the first hologram. We depict this process in Fig.~\ref{fig:setupOAM5} where we show an electron beam's recorded transverse profile as it propagates through the sorter. Here, we use electrons consisting of a superposition of states defined by OAM  values of $\pm 5$, i.e. $(\psi_{+5}(r,\phi)+\psi_{-5}(r,\phi))/\sqrt{2}$~\cite{shiloh:14}. Such a beam consists of a series of ten lobes that are equidistantly arranged along a ring shaped outline in the beam's transverse plane. The presence of these lobes is a direct result of the beam consisting of a superposition of two OAM components defined by the same magnitude, yet by opposite signs. More specifically, as they arise from the beam's OAM content, these lobes are additionally related to its constituent $\ell = \pm5$ electrons' transverse azimuthal phase profiles. The beam was generated using the phase mask whose TEM image along with an image of the beam transmitted through such a hologram can be found in the generator plane of the sorter found in Fig.~\ref{fig:setupOAM5}. After passing through the sorter's first hologram, these lobes rearrange themselves into a line since the beam's azimuthal phase variations have been unwounded along one of the Cartesian coordinates of its transverse plane. The beam propagates through the second hologram, shown in the corrector plane of Fig.~\ref{fig:setupOAM5}, in order to stabilize its propagation. The beam then propagates through a lens and is focused on a screen. This allows for the initial beam's OAM content to be readily analyzed as depicted in the final plane of Fig.~\ref{fig:setupOAM5}. 

\begin{figure}[t]
	\begin{center}
	\includegraphics[width=\columnwidth]{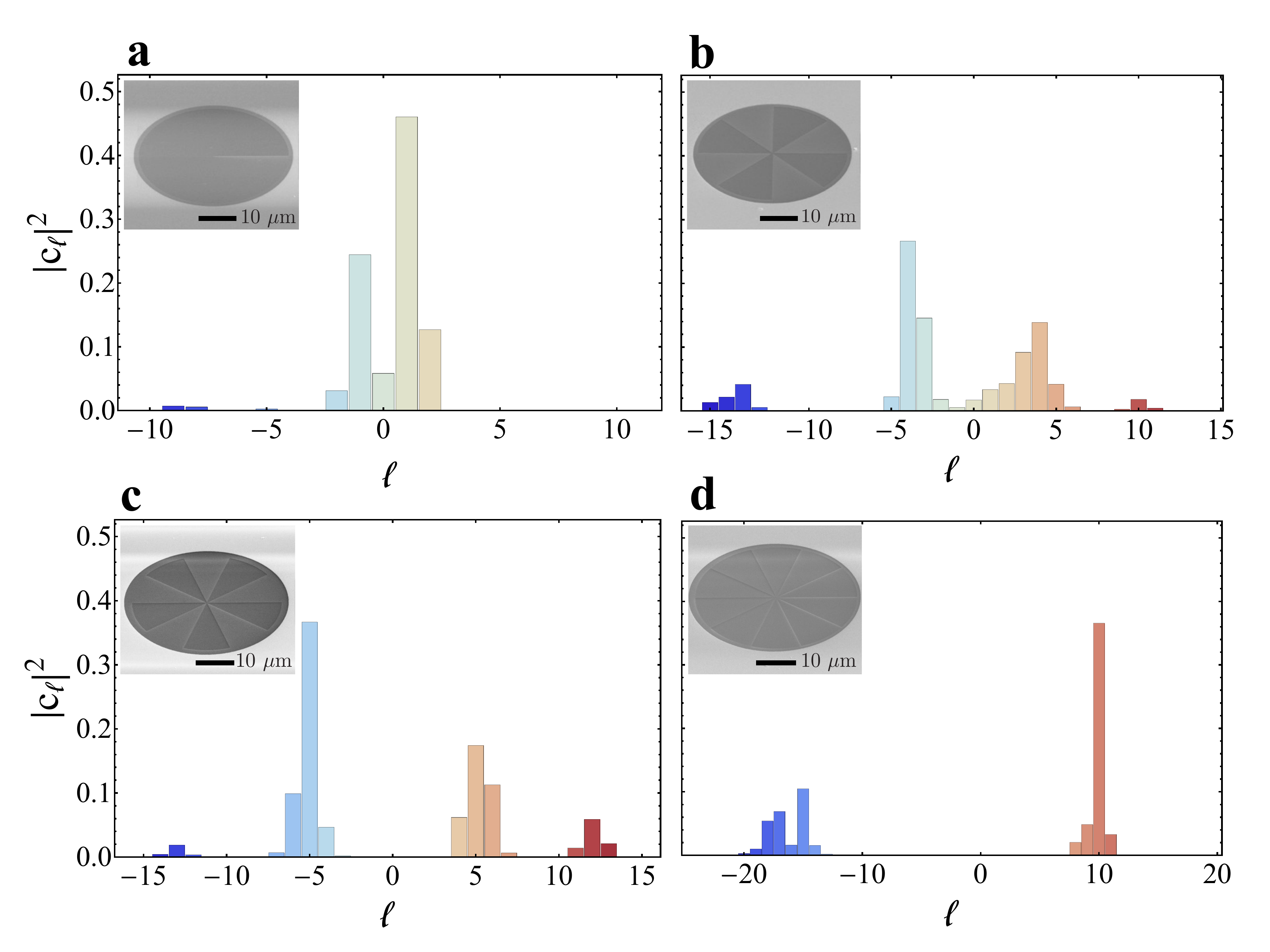}
	\caption[]{{\bf Experimental OAM spectra of various electron beams directly obtained from the sorter's output.} Spectrum of a beam consisting of electrons defined by: {\bf a} OAM of $+1$, $\psi_{+1}$, produced with a spiral phase plate; {\bf b} a superposition  of $\pm 4$ OAM states, $(\psi_{+4}+\psi_{-4})/\sqrt{2}$, generated by a phase mask;  {\bf c} a superposition of $\pm 5$ OAM states, $(\psi_{+5}+\psi_{-5})/\sqrt{2}$, generated by a phase mask; {\bf d} OAM of $+10$, $\psi_{+10}$, produced by a spiral phase plate. SEM images of the devices used to generate the analyzed electron beams are provided in the insets of their respective spetra.}
	\label{fig:OAMspectrum}
	\end{center}
\end{figure}

A more thorough calibration of the sorting apparatus was then performed by repeating the above process for wavefunctions carrying various values of OAM. The spectra resulting from this calibration have been background subtracted and deconvoluted using conventional spectroscopy methods (see Supplementary Materials). These results are displayed in Fig.~ \ref{fig:OAMspectrum} for the case of wavefunctions defined by single and superpositions of OAM states ranging from -10 to 10. Based on the various parameters defining the phase profile of the holograms used in the sorter, the cross-talk between components of the electrons' OAM spectra is expected to be below 20\%. However, due to fabrication and alignment imperfections in our apparatus, including the holograms generating the OAM carrying electrons, we observe higher values of cross-talk between OAM components. The values of the cross talk observed in the spectra shown in Fig.~ \ref{fig:OAMspectrum} were found to be {\bf a} 28\%,  {\bf b} 43\%, {\bf c} 39\%, and  {\bf d} 18\%. Though this may be the case, the sorter's performance is clearly observed through the distinct separation of the OAM states contained in the specific electron beams.

\begin{figure}[t]
	\begin{center}
	\includegraphics[width=\columnwidth]{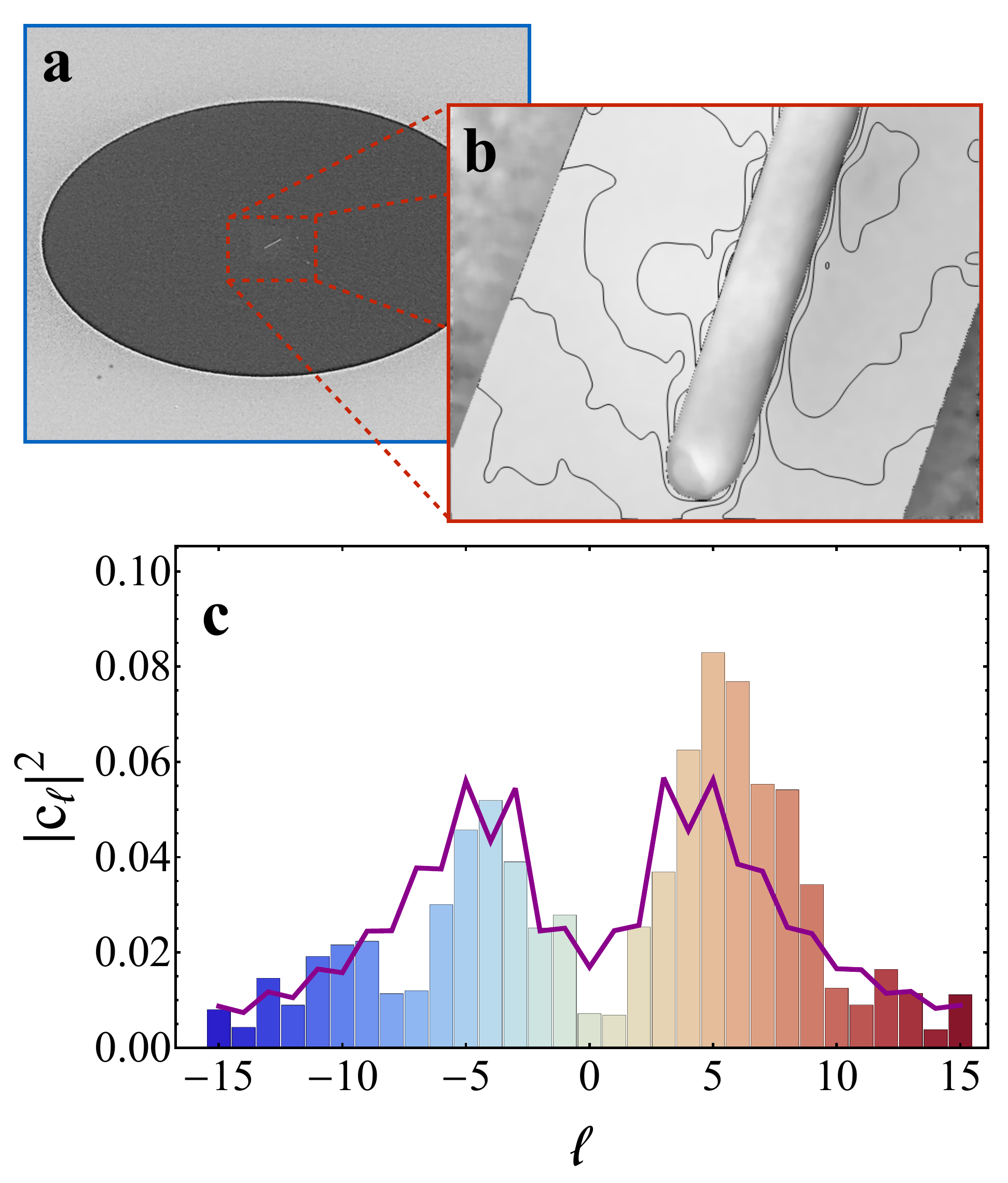}
	\caption[]{{\bf OAM spectrum of an electron beam having interacted with a magnetic dipole.} {\bf a} SEM image of the analyzed magnetic bar configurated as a dipole. The bar is defined by dimensions of 100~nm (thickness) by 200~nm (width) by 2.8~\textmu m (length).
{\bf b} Magnetic field lines of the magnetic bar. {\bf c} Expected (curve) and obtained (bars) OAM spectrum acquired by the electron beam upon interacting with the magnetic bar. The expected curve was calculated assuming that its magnetic field is saturated. Unlike the measurement performed in {\bf b}, the experimental data was obtained while the dipole was exposed to a field in the condenser plane of the electron microscope.}
	\label{fig:Dipole}
	\end{center}
\end{figure}

This sorter can be used to analyze magnetic structures affecting the OAM content of an electron beam. Here, we use our sorter to analyze the magnetic properties of a magnetized sample deposited using the method described in \cite{pozzi2016experimental}, a cobalt magnetic dipole in our case. To do so, the dipole has been positioned in the sorting apparatus in a way to replace the holograms that were previously generating the test wavefunctions. This magnetic structure consists of a single magnetic bar. A SEM image of the bar can be found in Fig.~{\ref{fig:Dipole}}-{\bf a}. Its magnetic configuration is namely characterized by a strong elongation enforcing the presence of a strong net magnetic dipole moment even after the magnetizing field is removed. Its remanence field was also characterized using electron holography and Lorentz imaging and is depicted in Fig.~\ref{fig:Dipole}-{\bf b}. 

The potential use of the sorter for this particular measurement arises from the fact that the term introduced by a magnetic dipole onto a passing Gaussian electron beam wavefunction can be written as
\begin{align}
	\label{eq:dipoleAction}
	g(r,\phi) = \exp{(i\,\chi(r)\sin{\phi})}
\end{align}
where $\chi (r) = (e \mu_0 \mathcal{M})/(h r)$, $h$ is the Planck constant, $\mu_0$ is the permeability of free space, and $\mathcal{M}$ is the dipole's magnetic moment. In order to observe the effect that such a term will have on an electron's OAM content, $g(r,\phi)$ must be Fourier-expanded in terms of $\phi$, namely $g(r,\phi)=\sum_{\ell=-\infty}^{\infty}c_\ell(r) \exp(i \ell \phi)$, where $c_\ell(r)$ are expansion coefficients. However, given that the resulting expansion terms carry a quantized azimuthal phase defined by $\ell$, then it follows that these components also carry OAM values of $\ell$. By default, the expansion coefficients $c_\ell$ correspond to the weight of each OAM component of a beam having been affected by the magnetic dipole. These coefficients can be found using the Jacobi-Anger expression and correspond to  $J_\ell(\chi(r))$, where $J_\ell$ is the $\ell^\text{th}$ order Bessel function of the first kind. This formulation of the added phase clearly depicts the imparted OAM content of the dipole onto the electron beam's wavefunction hence the ability to use the sorter to measure this magnetic dipole moment $\mathcal{M}$.

Though a more detailed analysis can be used to predict the OAM spectrum induced by the magnetic dipole, one can instead employ a simplified model to do so based on the apparatus' layout. The use of this model is justified by the truncation of the beam after passing through the first of the sorter's holograms. Such truncations result in lensing effects which will consequently displace electrons that were originally positioned at the hologram's cutoff radius $r_\text{max}$. As a result, these electrons will occupy a greater area of the beam following the truncation. Their profile will therefore also be spatially extended in the observed OAM spectrum. This will cause the general outline of the sorter's output to be predominantly associated with these electrons. Strictly speaking, the output spectrum of the sorter will be determined by its numerical apperture. Therefore, we may approximate the phase added by the dipole onto the beam as $\chi(r)\sin\phi\approx\chi(r_{\text{max}})\sin\phi$ which effectively removes the need to consider the radial dependence of the beam's OAM.  From this analysis, the relative probability of finding an electron of OAM $\ell\hbar$ in the resulting beam is given by the coefficient $\abs{c_\ell}^2=|{J_{\abs{\ell}}(\chi (r_\text{max}))}|^2$ which will also yield the beam's observed OAM distribution as a function of $\ell$. This analysis, in agreement with a more rigorous approach based on numerical calculations and also an analytical approach (see Supplementary Materials), reveals that the dominant decomposition coefficients of such an electron's spectrum will be attributed to $\ell$ values satisfying $|\ell|\approx\chi$.

Using the sorter, the OAM content of the wavefunction after its interaction with the dipole was recorded and is shown in Fig.~\ref{fig:Dipole}-{\bf c}. We find that the beam's OAM content is mostly distributed near $\ell=\pm 5$ thus implying that the dipole is defined by a $\chi$ value of approximately 5 radians. This value roughly translates to a magnetic dipole moment of $\mathcal{M}\approx 6.2 \times 10^9 \, \mu_B$, where $\mu_B$ is the Bohr magneton, and is in good agreement with our estimated value of the structure's saturated magnetic dipole moment of $6.7 \times 10^9 \, \mu_B$. The corresponding numerically simulated OAM spectrum based on these parameters is also included in Fig.~\ref{fig:Dipole}-{\bf c}. These numerical results were obtained using our simplified model where we assume that $\chi(r)\sin\phi\approx\chi(r_{\text{max}})\sin\phi$ and are in good agreement with data based on the saturation field of the wire. 

In comparison with other methods used to examine magnetic fields, c.f.~\cite{beleggia:10,saitoh:13,juchtmans:16}, the OAM sorter proves its effectiveness by readily providing the beam's OAM spectrum. For an identical number of detected electrons, this content is namely defined by an image showing twenty $|c_\ell|^2$ OAM coefficients yielding more information about a beam's phase than an image obtained using holographic methods. Moreover, such images do not allow a direct measurement of a sample's magnetic information. Instead, this quantity has to be extrapolated from the field in the dipole's proximity. In addition, the OAM sorter method does not rely on any phase wrapping or unwrapping methodology which simplifies a magnetic field's analysis. Further developments could also allow the improvement of this device and the possibility to exploit it in atomic scale measurements or in conjunction with scanning electron probes. Given that the sorter only requires the phase masks in two distinct planes, then its performance will become more effective if absorptive elements like phase holograms were substituted by structured electrostatic fields~\cite{beleggia:14}.

Our sorter method also possesses the following prospective extensions. On one hand, when using the sorter to analyze magnetic structures, the radial dependence of the phase added by such structures can be lowered by exposing them to a beam already carrying a known OAM value. This will cause the beam to have a maximal intensity at a certain radius $r_\ell$. Therefore, the majority of the beam's electrons having interacted with the structure acquire a phase whose radial dependence will be attributed to $r_\ell$. Much like how we approximated the phase acquired by electrons to be predominantly defined by the truncation radius of our apparatus $r_{\max}$, we could likewise add additional importance to electrons attributed to a radius of $r_\ell$. Such an approximation could provide additional simplifications that are needed for matching the outcome of these magnetic measurements with theory. On the other hand, a minor modification to the sorter's schematics can be brought to sort electron modes in a different mutually unbiased basis, namely the so-called angular basis~\cite{mirhosseini:13}. Performing an additional set of measurements of an electron's phase in this basis could provide additional information, e.g. regarding the phase of the OAM expansion coefficients $c_\ell$~\cite{bent:15}, concerning the interaction of a magnetic field with electron beams. 


\vspace{1cm}
\noindent\textbf{Supplementary Materials} Materials and Methods, Supplementary Text, and Fig.~\ref{fig:processing}.
\vspace{1 EM}

\noindent\textbf{Acknowledgments}
\noindent V.G. acknowledges the support of the Alexander von Humboldt Foundation. H.L., F.B, R.W.B. and E.K. acknowledge the support of the Canada Research Chairs (CRC) and Canada Excellence Research Chairs (CERC) Program.
\vspace{1 EM}

\noindent\textbf{Author Information}
\noindent The authors declare no competing financial interests. Correspondence and requests for materials should be addressed to E.K. (ekarimi@uottawa.ca) or V.G. (vincenzo.grillo@unimore.it).

\clearpage
\setcounter{figure}{0} \renewcommand{\thefigure}{S\arabic{figure}}
\setcounter{table}{0} \renewcommand{\thetable}{S\arabic{table}}
\setcounter{section}{0} \renewcommand{\thesection}{S\arabic{section}}
\setcounter{equation}{0} \renewcommand{\theequation}{S\arabic{equation}}
\onecolumngrid
\section*{{\Large Supplementary Materials for}\\ Measuring an electron beam's orbital angular momentum spectrum}

\section*{\underline{\large{Part 1:}} Devices to generate test wavefunctions}
To generate OAM carrying electrons, a $\exp{(i\,\alpha(\phi))}$ term must be added to the formulation of their wavefunctions, where $\alpha(\phi)$ is a function depending on the azimuthal coordinate ($\phi \in [0,2\pi]$). To do so, we employ phase masks which will introduce some losses to the intensity of the electron beam which can be attributed to absorption. We modify the term imparted to the wavefunctions to $\exp{(i\,\alpha(\phi)-a(\phi))}$, where $a$ is a positive constant accounting for any absorption related effects.

\subsection{Two-level masks}

The phase designed to be added by a two level mask is defined by
\begin{align}
	\alpha(\phi) = \begin{cases} 0 & \quad0<\mod(n\phi,2\pi)<\pi \\ \delta_0 & \quad\pi<\mod(n\phi,2\pi)<2\pi\end{cases}.
\end{align}
where $\mod(x,b)$ yields the remainder of the division of $x$ by $b$ and $n$ is an integer. It thus follows that such a plate's influence on an electron's wavefunction is provided by the terms
\begin{align}
	\widetilde\psi(\phi) = \begin{cases} 1 & 0<\mod(n\phi,2\pi)<\pi \\ \exp(-a+i\,\delta_0)& \pi<\mod(n\phi,2\pi)<2\pi\end{cases}.
\end{align}
To obtain the OAM components of a beam generated by such a mask, we must find the expansion coefficients of this term's Fourier series expanded in $\phi$, namely
\begin{eqnarray}
	c_\ell &=& \frac{1}{2\pi}\int_0^{2\pi} \widetilde{\psi}(\phi) \exp(i\,\ell \phi)\,d\phi\\[7.5pt]
	&=& 
	\begin{cases}
		\frac{n}{\ell\pi}[-1+\exp(-a+i\,\delta_0)]& \ell=nm \\
		0& \ell\neq nm \\
		\frac{1}{2}[1+\exp(-a+i\,\delta_0)]& \ell=0
	\end{cases}
\end{eqnarray}
where $m$ is an odd integer. For non-absorptive masks, i.e. $a=0$, these coefficients become
\begin{eqnarray}
	c_\ell 	&=& 
	\begin{cases}
		\frac{2n}{\ell\pi}(\exp(i\,\delta_0/2)\sin(\delta_0/2))& \ell=nm \\
		0& \ell\neq nm \\
		\exp(i\,\delta_0/2)\cos(\delta_0/2)& \ell=0
	\end{cases}.
\end{eqnarray}
Due to their well-defined periodicity, these  holograms have very strong selection rules.

\subsection{Spiral masks}
These masks are designed to add a phase of $\alpha(\phi)=(\delta_0 \mod(n\phi,2\pi))/2\pi$ to electron wavefunctions. As in the case of two-level masks, we use the corresponding Fourier coefficients to determine the OAM content acquired by electrons upon propagation through this type of element. These coefficients are provided by
\begin{align}
	|c_\ell|=
	\begin{cases}
		\frac{\sin\left(\frac{2\pi\ell}{n}+\delta_0\right)}{\ell+\delta_0/2\pi},&  \ell = nm\\
		0,&  \ell \neq nm
	\end{cases}	
\end{align}
where $m$ is a positive integer. For $\delta_0=2\pi$, the hologram produces a vortex beam carrying OAM of $\ell=n$.

\section*{\underline{\large{Part 2:}} Sorter and corrector holograms}
The required ``unwrapping'' process can be achieved by means of a log-polar coordinate transformation through the use of two coupled diffractive holograms. However, due to some differences between the optical sorter and our electron microscope-based apparatus, we must resort to bringing slight adaptive corrections to the phase grating of these holograms.  On one hand, the phase corresponding to the first one is modified to
\begin{align}
	\Lambda_1 = \phi_0 \sign{\left(\sin{\left(2\pi \, a\left[y\arctan{\left(\frac{y}{x}\right)}+x\ln{\left(\frac{\sqrt{x^2+y^2}}{b}\right)}+x\right]\right)}\right)}
\end{align}
where $a$ and $b$ are two parameters scaled to optimize the method's experimental efficiency while $\sign$ denotes the sign function. On the other hand, the added phase associated with the second hologram becomes
\begin{align}
	\Lambda_2 = \phi_1 \sign{\left(\sin{\left(2\pi \, a \, b \, \exp{\left(-2\pi \frac{u}{a}\right)}\cos{\left(2\pi\frac{v}{a}\right)}\right)}+2\pi \,c\,v\right)},
\end{align}
where $u = -a\ln{\left(\sqrt{x^2+y^2}/b\right)}$, $v = a \arctan{(y/x)}$, and $c$ consists of an additional scaling parameter. For our sorter, we used parameters of $a=2$, $b=0.01$, and $c=0.6$. The factors $\phi_0$ and $\phi_1$ in the phase expressions of both elements should take values of $\pi$ in order to maximize their respective diffraction efficiencies. Moreover, in order to adapt this method to a transmission electron microscope (TEM), supplementary sets of lenses and apertures were also added.

\section*{\underline{\large{Part 3:}} OAM spectrum processing}
\begin{figure}[h]
\begin{center}
\includegraphics[width=\columnwidth]{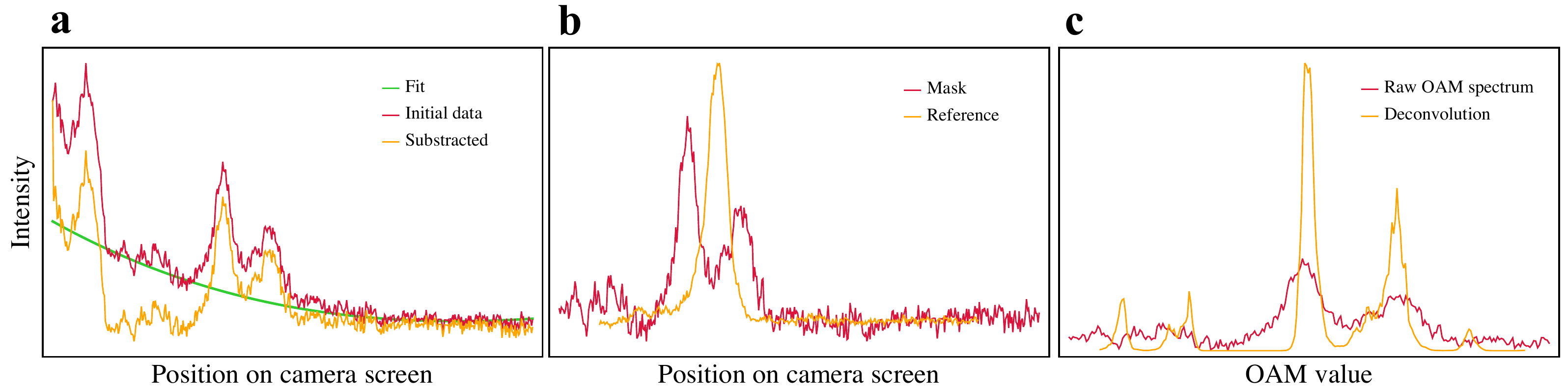}
\caption[]{{\bf Obtained OAM spectra at various stages of its processing.} {\bf a} Raw data obtained from the sorter's output along with the fitted background and the background subtracted data. {\bf b} Background subtracted data along with the reference zero-OAM electron beam used for the sorter's calibration. {\bf c} Raw calibrated OAM spectrum and its deconvoluted version.}
\label{fig:processing}
\end{center}
\end{figure}

Experimental OAM spectra obtained from the sorter were analyzed using conventional spectroscopy techniques. Background noise is first subtracted from the obtained signal by fitting it to a third order polynomial and then subtracting the fit from the raw data as shown in Fig.~\ref{fig:processing}-{\bf a}. The OAM scale of the spectra is then calibrated using data obtained from a reference beam. In our case, this reference consisted of a superposition of $\pm4$ OAM states and a reference beam carrying no OAM, both of which are depicted in Fig.~\ref{fig:processing}-{\bf b}. The calibrated signal is then deconvoluted using the maximum entropy method after clipping negative values in the spectra. The effect of the employed deconvolution algorithm is shown in Fig.~\ref{fig:processing}-{\bf c}. Finally, a binning procedure is conducted by assembling all pixels between values of $\ell-0.5$ and $\ell+0.5$ from which we obtain discretized OAM spectra. Optimizations of the binning offsets are also performed in order position the spectra's maxima in the centre of their respective bins.

\section*{\underline{\large{Part 4:}} Analysis of the interaction between the dipole and electrons}
We can obtain the expansion coefficient $c_\ell$ attributed to the phase added by a magnetic dipole onto an electron beam by performing a Fourier transform of the term, $g$ as introduced in Eq.~(\ref{eq:dipoleAction}), added by the dipole onto the wavefunction of the electrons. This transformation must be performed in terms of the Cartesian coordinates $(u,v)$ of the plane in which the beam is unwrapped. It therefore follows that these coefficients are given by
\begin{align}
	c_{p,\ell}=\int_{-\infty}^{+\infty}\int_{-\infty}^{+\infty} \exp(i\,\chi (r(u))\sin\phi(v)) \exp(i\,\ell v)\exp(i \,u p) \,du\,dv
\end{align}
Given that our transformed coordinates are defined by $u = -a\ln(r/b)$ and $v=a \phi$, we perform according variable substitutions in order to obtain an expression for the coefficients in terms of polar coordinates. For simplicity, we assume that $a$ and $b$ are equal to 1 and obtain
\begin{eqnarray}
	c_{p,\ell} &=&\int_{0}^{2\pi}\int_{0}^{\infty} \exp(i\,\chi (r)\sin\phi)  \exp(i\,\ell \phi)\exp(-i \,\ln(r) p) \,dr\,d\phi 
\end{eqnarray}
In the case where an intermediate aperture is added to sorter's configuration, we may model this addition by modulating the above integrand by a Gaussian function or more precisely a radial step-function $W(r)$. In the case where this aperture is sufficiently narrow, we may consider that the main coefficients in the OAM spectrum are for $p$ values satisfying $p\approx 0$ hence allowing us to neglect any other values of $p$. Based on these considerations, we may further simplify our expansion coefficients to
\begin{eqnarray}
	c_{\ell} \approx c_{0,\ell}&=&\int_{0}^{2\pi}\int_{0}^{\infty} \exp(i\,\chi (r)\sin\phi)  \exp(i\,\ell \phi)W(r) \,dr\,d\phi .
\end{eqnarray}
We then expand $\exp(i\,\chi (r)\sin\phi)$ using the Jacobi-Anger identity. Due to the orthonormality between the exponentials in the resulting expression, we may further simplify the expansion coefficient to 
\begin{eqnarray}
	c_{0,\ell}&=&\int_{0}^{\infty} J_\ell(\chi(r)) W(r) \,dr \approx \int_{r_0-\sigma}^{r_0+\sigma} J_\ell(\chi(r))\,dr
\end{eqnarray}
where the bounds of integration $r_0-\sigma$ and $r_0+\sigma$ are used to approximate the influence of the modulating apperture function on the expansion coefficients. From this expression, we observe that $c_{0,\ell}\approx 0$ when $\chi(r)\ll \ell$. Likewise, in the case where $\chi(r)\gg \ell$, the integrand oscillates rapidly and integration over these larger $r$ values will likewise yield negligible contributions. Therefore, these $c_{0,\ell}$ coefficients will only hold significant values over a range where $\chi(r)$ is near $|\ell|$. 
\end{document}